\documentclass[twocolumn]{aastex631}

\shorttitle{ML for Giant Impact Simulations}
\shortauthors{Lammers et al.}

\graphicspath{{./}{}}
\usepackage{amsmath}
\usepackage{comment}
\begin{document}

\newcommand{\paren}[1]{\left(#1\right)}
\newcommand{\fracbrac}[2]{\paren{\frac{#1}{#2}}}
\newcommand{\pp}{\phantom{$-$}}

\title{Accelerating Giant Impact Simulations with Machine Learning}

%\correspondingauthor{Caleb Lammers}
%\email{caleb.lammers@princeton.edu}

\author[0000-0001-9985-0643]{Caleb Lammers}
\affiliation{Department of Astrophysical Sciences, Princeton University, 4 Ivy Lane, Princeton, NJ 08544, USA}

\author[0000-0002-6458-3423]{Miles Cranmer}
\affiliation{Department of Applied Mathematics and Theoretical Physics, University of Cambridge, Wilberforce Road, Cambridge, CB3 0WA, UK}
\affiliation{Institute of Astronomy, University of Cambridge, Madingley Road, Cambridge, CB3 0HA, UK}
\affiliation{Kavli Institute for Cosmology, University of Cambridge, Madingley Road, Cambridge, CB3 0HA, UK}

\author[0000-0002-1032-0783]{Sam Hadden}
\affiliation{Canadian Institute for Theoretical Astrophysics, University of Toronto, 60 St.\ George Street, Toronto, ON M5S 3H8, Canada}

\author[0000-0002-1068-160X]{Shirley Ho}
\affiliation{Center for Computational Astrophysics, Flatiron Institute, New
York, NY 10010, USA}
\affiliation{Department of Astrophysical Sciences, Princeton University, 4 Ivy Lane, Princeton, NJ 08544, USA}
\affiliation{Department of Physics \& Center for Data Science, New York University, 726 Broadway, New York, NY 10003, USA}

\author[0000-0002-8659-3729]{Norman Murray}
\affiliation{Canadian Institute for Theoretical Astrophysics, University of Toronto, 60 St.\ George Street, Toronto, ON M5S 3H8, Canada}
\affiliation{Department of Physics, University of Toronto, 60 St.\ George Street, Toronto, ON M5S 1A7, Canada}

\author[0000-0002-9908-8705]{Daniel Tamayo}
\affiliation{Department of Physics, Harvey Mudd College, Claremont, CA 91711, USA}

\begin{abstract}

Constraining planet formation models based on the observed exoplanet population requires generating large samples of synthetic planetary systems, which can be computationally prohibitive. A significant bottleneck is simulating the giant impact phase, during which planetary embryos evolve gravitationally and combine to form planets, which may themselves experience later collisions. To accelerate giant impact simulations, we present a machine learning (ML) approach to predicting collisional outcomes in multiplanet systems. Trained on more than 500,000 $N$-body simulations of three-planet systems, we develop an ML model that can accurately predict which two planets will experience a collision, along with the state of the post-collision planets, from a short integration of the system's initial conditions. Our model greatly improves on non-ML baselines that rely on metrics from dynamics theory, which struggle to accurately predict which pair of planets will experience a collision. By combining with a model for predicting long-term stability, we create an ML-based giant impact emulator, which can predict the outcomes of giant impact simulations with reasonable accuracy and a speedup of up to four orders of magnitude. We expect our model to enable analyses that would not otherwise be computationally feasible. As such, we release our training code, along with an easy-to-use API for our collision outcome model and giant impact emulator.\footnote{\url{https://github.com/dtamayo/spock}}

\end{abstract}


\section{Introduction}
\label{sec:intro}

The nebular hypothesis, initially proposed by Immanuel Kant \citep{Kant1755} and later built upon by Pierre Laplace \citep{Laplace1796}, remains the leading explanation for the formation of planetary systems. In the modern picture, terrestrial planets form out of a protoplanetary disk before experiencing a phase of giant impacts in which planets grow via collisions (see, e.g., \citealt{Lissauer1993}). Giant impacts are therefore believed to play an important role in shaping the orbital configuration of planetary systems, and there has been significant effort to simulate this process \citep[e.g.,][]{Kokubo&Ida1998, Chambers&Wetherill1998, Agnor1999, Chambers2001, Raymond2005, Kokubo2006, Raymond2008, Hansen&Murray2012}.

Interest in terrestrial planet formation has been re-invigorated by the discovery, due to  NASA's {\it Kepler} mission, that compact multiplanet systems are remarkably common in the galactic neighbourhood \citep{Borucki2011, Lissauer2011, Fressin2013, Fabrycky2014, Zhu2018}. Planets in these systems typically possess short periods, sub-Neptune radii and masses, low eccentricities, and low inclinations. Motivated by the tight dynamical spacing of planets in observed systems \citep{Pu&Wu2015, Volk&Gladman2015} and the inability of these systems to host additional intervening planets \citep{Fang&Margot2013, Obertas2023}, it is believed that observed systems are shaped by dynamical instabilities arising over their gigayear lifetimes. In general, close encounters between planets can result in planet-planet collisions, planet-star collisions, or ejections. However, planets in typical transiting multiplanet systems orbit with semi-major axes for which dynamical instabilities nearly always result in planet-planet collisions.

Numerical simulations of the giant impact phase have been successful in reproducing many properties of the observed population of multiplanet systems \citep[e.g.,][]{Hansen&Murray2013, Izidoro2017, Poon2020, Goldberg&Batygin2022, Lammers2023, Ghosh&Chatterjee2024}. These simulations typically begin with a collection of overly packed planets\footnote{Low-mass bodies in $N$-body simulations are sometimes referred to as ``planetary embryos'' or ``protoplanets.'' For simplicity, throughout this work, we will refer to the bodies in our $N$-body as ``planets'' regardless of their mass.} and evolve forward in time using a hybrid integrator (e.g., \texttt{mercury6} \citealt{Chambers1999} or \texttt{MERCURIUS} \citealt{Rein2019}). Hybrid integrators make long-term integrations computationally feasible by using a rapid Wisdom-Holman integrator \citep{Wisdom&Holman1991} when planets are widely separated, switching to a high-precision integrator only when it is necessary to resolve a close encounter. Before reaching the maximum integration time, systems in these simulations continually destabilize, resulting in collisions between planets (which are typically treated as perfect inelastic mergers). Due to the chaotic nature of multiplanet systems, outcomes of such numerical experiments are only meaningful in a statistical sense, necessitating simulations of many realizations with different initial conditions.

Efforts to numerically model this process are limited by the computational cost of running large samples of long-term $N$-body simulations. In recent years, machine learning (ML) techniques have been successfully applied to accelerate $N$-body simulations, both in the cosmological \citep{He2019, Jamieson2023} and planetary \citep{Tamayo2020, Cranmer2021} contexts. In particular, \citet{Tamayo2020} (hereafter, SPOCKI) presented an ML model to predict whether a compact planetary system is stable over $10^9$ orbits (here, and throughout the paper, the ``orbits'' unit refers to the initial orbital period of the innermost planet), based on summary statistics calculated over a short $10^4$ orbit integration. Extending this work, \citet{Cranmer2021} (hereafter, SPOCKII) developed a model that, when provided the raw orbital elements of the planets from a $10^4$ orbit integration, can accurately predict the time at which a compact planetary system will destabilize. These tools have enabled investigations into the long-term stability of planetary systems that would not otherwise be computationally feasible \citep[e.g.,][]{Tamayo2021, Yee2021, Bailey&Fabrycky2022, Tejada2022, Obertas2023, Sobski&Millholland2023}. The development of SPOCKII was also motivated by the possibility of accelerating giant impact simulations. Towards this end, here we present an ML-based approach to predicting the outcomes of planet-planet collisions in $N$-body simulations.

This paper is organized as follows: Section~\ref{sec:MLmodel} describes our training set and machine learning model. We evaluate the performance of our model and compare it with non-ML approaches in Section~\ref{sec:results}. Discussion and conclusions are presented in Section~\ref{sec:discussion}.

\section{Machine learning model}
\label{sec:MLmodel}

\subsection{Training dataset}
\label{sec:dataset}

To train our ML model, we require a large collection of $N$-body simulations in which two planets experience a physical collision. Unfortunately, the SPOCKI training dataset (which was also used in SPOCKII) is unsuitable for this purpose due to choices made to tailor the dataset for predicting long-term stability (e.g., stopping $N$-body integrations when the Hill spheres of two planets intersect). As a result, for this work, we generate a new training set of $N$-body simulations. Motivated by the intention that our model will often be used in tandem with SPOCKII, we sample initial conditions that span a similar region of parameter space.

As in the SPOCK training set, we restrict our dataset to systems of three planets. This simplification is motivated by the finding that there is a qualitative change in the instability behavior between two- and three-planet systems, but little change when additional planets are included in the system \citep[e.g.,][]{Chambers1996, Petit2020}. Theoretically, this is believed to be due to the importance of mean motion resonances (MMRs) between adjacent planets, and the secondary resonances they generate, in driving multiplanet systems to destabilize \citep{Quillen2011, Petit2020, Lammers2024}. Because three-body interactions are believed to drive the instability process, we expect that restricting the training set to three-planet systems is also a reasonable assumption for the task of predicting collisional outcomes.

We draw planet-to-star mass ratios for each of the three planets log-uniformly from the range [$10^{-7}$, $10^{-4}$], spanning approximately from the mass ratio of small planetary embryos to larger than that of Neptune. Orbital inclinations were sampled log-uniformly from the narrow range of [$10^{-3}$\,rad, $0.3$\,rad] to reflect that most observed multiplanet systems are nearly coplanar \citep{Fang&Margot2012, Fabrycky2014, HeM2019}. Nonetheless, this choice allows for mutual inclinations of up to $34^\circ$ in the training set. Initial orbital angles (i.e., longitude of the ascending node, longitude of pericenter, and true longitude) were drawn uniformly from [$0$, $2\pi$]. Eccentricities for each planet were drawn log-uniformly from [$10^{-3}$, $e_\mathrm{cross}$], where $e_\mathrm{cross}$ is the eccentricity at which the orbit of the planet crosses that of its neighbor:
\begin{equation}
    e_\mathrm{cross} = \frac{a_{i+1} - a_i}{a_{i+1}}
\end{equation}
where $i\,{=}\,1$ for the innermost planet and $i\,{=}\,2$ for the outermost planet. For the middle planet, we take $e_\mathrm{cross}$ to be the minimum eccentricity required to cross either the orbit of the inner planet or the outer planet (i.e., $e_\mathrm{cross}\,=\mathrm{min}\{(a_2\,{-}\,a_1)/a_2,\,(a_3\,{-}\,a_2)/a_3\}$).

The spacing of the planets was chosen so that systems span a range of initial separations, but are nonetheless likely to destabilize. It has been shown that instability times in multiplanet systems scale with the spacing unit $S_i\,{=}\,(a_{i+1}\,-\,a_i)/(a_{i+1}\mu^{1/4})$ where $\mu\,=\,(m_i + m_{i+1})/M_\ast$ \citep{Petit2020, Lammers2024}. Leveraging this, we arbitrarily place the innermost planet at $a_1\,{=}\,1.00$ and sample $S_1$ uniformly from [$0.0$, $7.5$] and then $S_2$ uniformly from [$0.0$, $7.5$], to determine the semi-major axes of the outer two planets. Planet radii were assigned using the mass-radius relationship $(M/M_\oplus)\,{=}\,2.7(R/R_\oplus)^{1.3}$ from \citet{Wolfgang2016}. Note, however, that we must assume a typical value of $a_1$ to convert planet radii from units of $R_\oplus$ to units in which $a_1\,{=}\,1.0$; we adopt a nominal value of $a_1\,{=}\,0.1$\,AU, interior to which the prevalence of super-Earths declines. We caution that the predictions of the model may be inaccurate for systems of planets with densities that differ drastically from the adopted mass-radius relationship.

\begin{figure*}
\centering
\includegraphics[width=\textwidth]{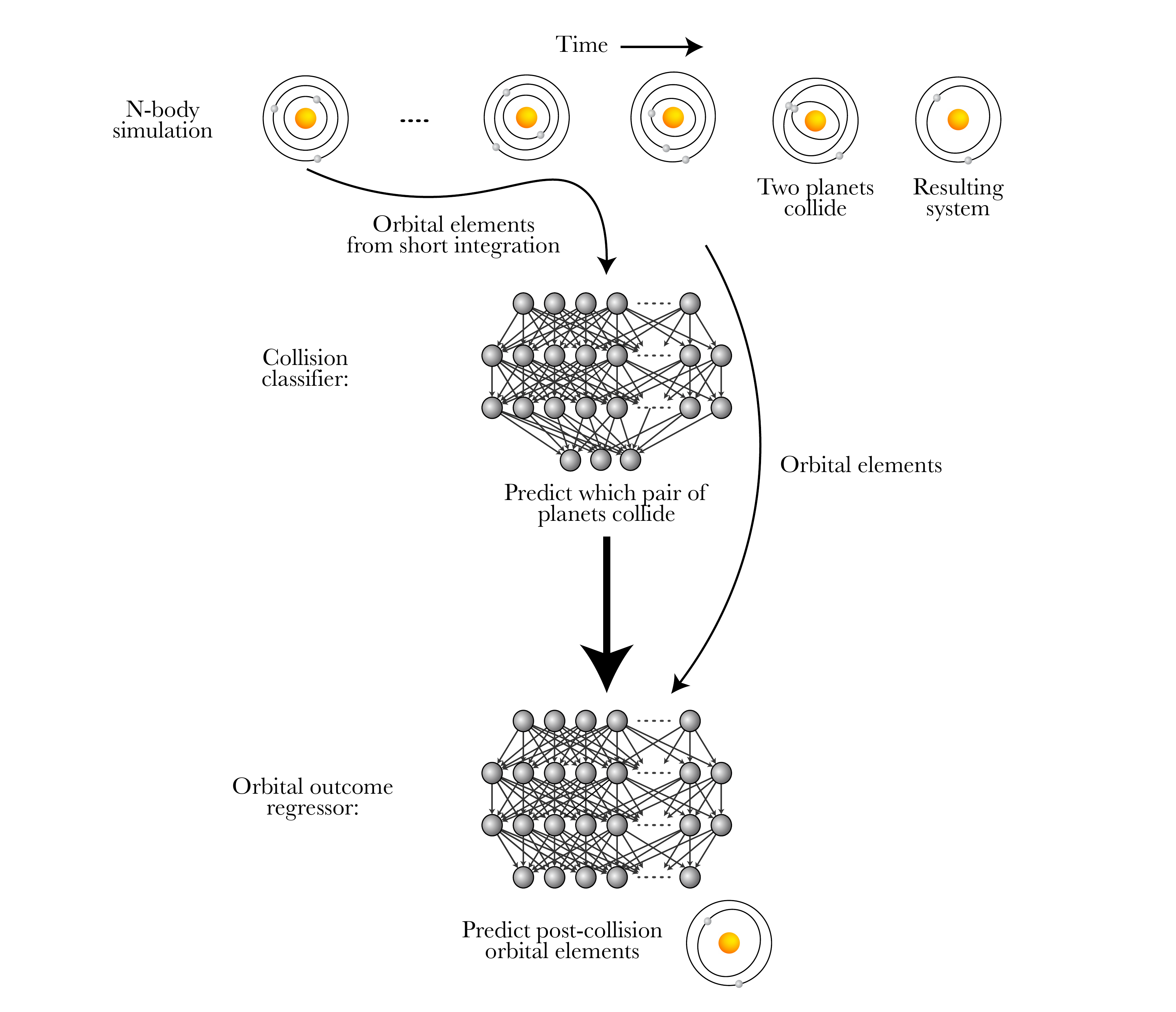}
\caption{Schematic of our machine learning model. The collision classifier takes the mean and standard deviation of the three planets' orbital elements from a short $N$-body integration as input and predicts which pair of planets will collide (in the form of three collision probabilities). The orbital outcome regressor takes the orbital elements, as well as a choice for which two planets to combine, and predicts the orbital elements of the two resulting post-collision planets.}
\label{fig:NN_schematic}
\end{figure*}

$N$-body integrations were performed using the \texttt{MERCURIUS} integrator \citep{Rein2019} from the open-source \texttt{REBOUND} code. \texttt{MERCURIUS} is a hybrid integrator, which uses a fixed-timestep Wisdom-Holman integrator (\texttt{WHFast}; \citealt{Wisdom&Holman1991, Rein&Tamayo2015}) when the planets are widely separated and switches smoothly to a high-order adaptive integrator (\texttt{IAS15}; \citealt{Rein&Spiegel2015}) when planets come within three Hill radii of each other. As recommended by \citet{Wisdom2015}, we adopt a timestep of $T_p/20$, where $T_p$ is the minimum perihelion passage timescale of the three planets. For consistency across systems, before integrating, we align the $z$-axis of our coordinate system with the angular momentum vector of the three-planet system.

As in previous works, we treat collisions between planets as perfect inelastic collisions in which mass and momentum are conserved.\footnote{Adopting a more sophisticated merger prescription does not meaningfully impact the resulting orbital configurations \citep{Poon2020, Esteves2022}.} We integrate systems for $10^7$ orbits of the innermost planet, stopping the simulation prematurely if two planets merge. Our choice of stopping time is motivated by the typical time required for a physical merger to occur \citep{Rice2018} and makes it computationally feasible to generate a training set of $500$,$000+$ systems. Note that SPOCKI and SPOCKII were trained to predict stability on timescales up to $10^9$ orbits. However, whereas there is a clear advantage to increasing the integration time for stability predictions, there is less of an advantage for our task; it is unlikely that mergers that occur in systems which take longer to destabilize differ meaningfully from mergers experienced by systems that destabilize more quickly.

While constructing the training dataset, we reject systems that experienced a merger in less than $10^4$ orbits (collision outcome predictions are not needed for such systems because short integrations are computationally inexpensive) or survive $10^7$ orbits without a merger. Restricting the training set to systems that experience mergers allows us to train a model that focuses exclusively on predicting collisional outcomes, leaving aside the challenge of predicting stability, which was the focus of SPOCKI and SPOCKII. After removing $1$,$540$ systems that experience an ejection (i.e., systems in which a planet is scattered to $a\,{>}\,50$\,$a_1$), the final training set consists of $517$,$016$ systems. Note that the distributions of initial orbital elements in the training set are affected by the requirement that systems must experience a merger between $10^4$ and $10^7$ orbits (e.g., compact, eccentric systems typically experience a merger in less than $10^4$ orbits and are thereby excluded).

\subsection{Model description}
\label{sec:architecture}

We break up the task of predicting planet-planet collision outcomes into two subproblems: (1) predicting which pair of planets will experience a collision and (2) predicting the orbital configuration of the post-collision system. The former is a classification problem with three possibilities: planet $1$ collides with planet $2$ ($43.3$\,\% of systems in the training set), planet $2$ collides with planet $3$ ($37.4$\,\% of systems), or planet $1$ collides with planet $3$ ($19.3$\,\% of systems). Predicting the orbital elements of the post-collision system, on the other hand, is a regression problem with multiple outputs. We explored different ML model architectures for tackling these tasks, but we found a shallow multi-layer perceptron (MLP) model to work well for both. A schematic of our ML model, encompassing both collision classifier and orbital outcome regressor parts, is shown in Fig.~\ref{fig:NN_schematic}.

\begin{figure*}
\centering
\includegraphics[width=\textwidth]{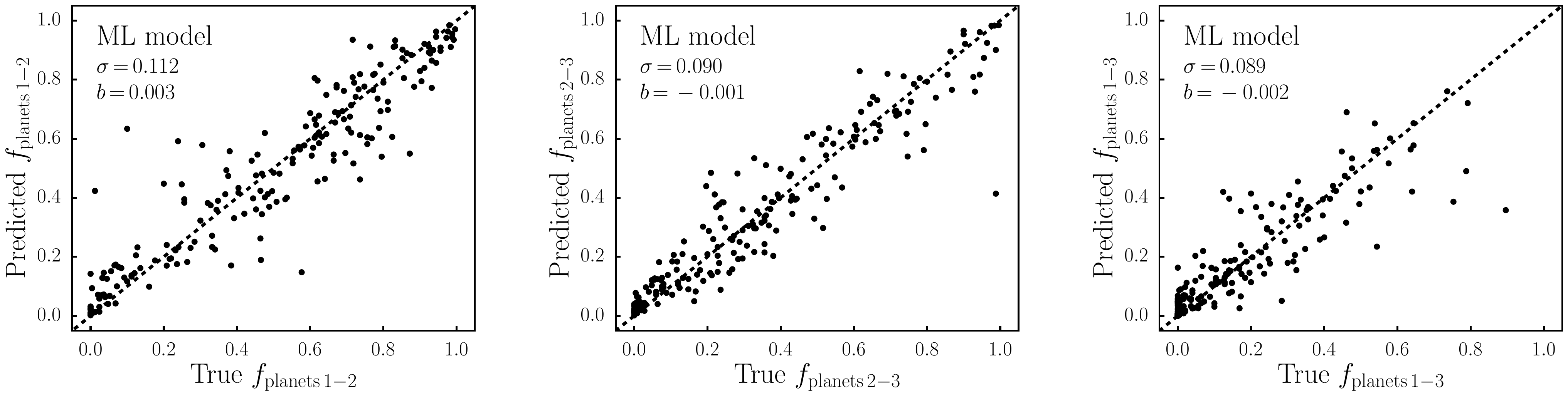}
\caption{Performance of the collision classifier model on $200$ random validation systems. The probabilities of a collision occurring between planets $1$\,--\,$2$, $2$\,--\,$3$, and $1$\,--\,$3$, as predicted by the ML model, are plotted against the true fraction of collisions that occur between the planet pairs, determined by performing $250$ shadow integrations of each system. The model accurately predicts collision probabilities with a scatter of ${\sim}\,10$\,\% and very little bias about the true fractions ($\sigma$ is the root-mean-squared error and $b$ is the mean of the residuals). For comparison, all baseline models we considered performed poorly (see Appendix~\ref{sec:classifer_baseline}).}
\label{fig:classification_accuracy}
\end{figure*}

To address the classification problem, we construct a model that, when provided information about the system as input, outputs three numbers: the probability of a collision occurring between planets $1$\,--\,$2$, planets $2$\,--\,$3$, and planets $1$\,--\,$3$. Motivated by previous works (\citealt{Tamayo2016}; SPOCKI; SPOCKII), as input, we provide the MLP model with the mean and standard deviation of the orbital elements of the three planets over a short $10^4$ orbit integration of the system. The time required to perform a $10^4$ orbit integration is approximately equal to the inference time of the model, so there is no computational gain to generating inputs with a shorter integration. The classification MLP was trained to minimize binary cross-entropy loss with the Adaptive Moment Estimation Optimizer (ADAM; \citealt{Kingma&Ba2014}) for $1$,$000$,$000$ optimizer steps. The hyperparameters of the model (network depth\,${=}\,1$ hidden layer, number of hidden nodes\,${=}\,30$, learning rate\,${=}\,7\,{\times}\,10^{-4}$, weight decay\,${=}\,1\,{\times}\,10^{-4}$, and mini-batch size\,${=}\,1$,$000$) were determined prior to training with Bayesian optimization on a pared-down version of the training set, and the final model was trained on $80$\,\% of the full training set ($413$,$612$ systems). A full description of the model architecture and training process can be found in Appendix~\ref{sec:classifer_details}.

To address the regression problem, we require a model that can predict the post-collision orbital configuration of the system, given a short integration of the system and a choice of which two planets are to collide. Although only two of the three planets are involved in the collision, we find that the state of the third planet can also change substantially during the instability phase. As a result, we require a model that can predict the orbital parameters of the newly merged planet, along with those of the surviving planet.

In practice, we found it challenging to predict the orbital angles (longitude of the ascending node, longitude of pericenter, and true longitude) of the two post-collision planets. Further investigation reveals that, in integrations of the synthetic systems where the initial conditions have been slightly perturbed (see Section~\ref{sec:validation_set}), the orbital angles of the planets typically span from approximately $[0,\,2\pi]$ across the set of perturbed systems. As a result, we trained the regression model to predict only semi-major axes, eccentricities, and inclinations, and when the orbital angles are needed for a particular application of the model (e.g., replacing three planets in a synthetic planetary system with two post-collision planets), we simply draw them randomly from $[0,\,2\pi]$. It may prove valuable to work towards a model that can predict dynamically relevant combinations of the orbital angles (e.g., mutual inclination) in the future.

As input, we provide the regression model with the mean and standard deviation of the orbital elements of the three planets, ordered to specify which two planets are involved in the collision. The regression MLP was trained using a mean-squared-error loss function with a correction term to account for the upper limits on the orbital element outputs (see Appendix~\ref{sec:regressor_details} for more details). The model was trained on $80$\,\% of the full training set using the ADAM optimizer for a total of $1$,$000$,$000$ optimizer steps. Like the classification model, hyperparameters (network depth\,${=}\,1$ hidden layer, number of hidden nodes\,${=}\,60$, learning rate\,${=}\,7\,{\times}\,10^{-4}$, weight decay\,${=}\,1\,{\times}\,10^{-4}$, and mini-batch size\,${=}\,3$,$000$) were determined with Bayesian optimization performed on a subset of the training set.

\subsection{Giant impact emulator}
\label{sec:formationpreds}

\begin{figure*}
\centering
\includegraphics[width=\textwidth]{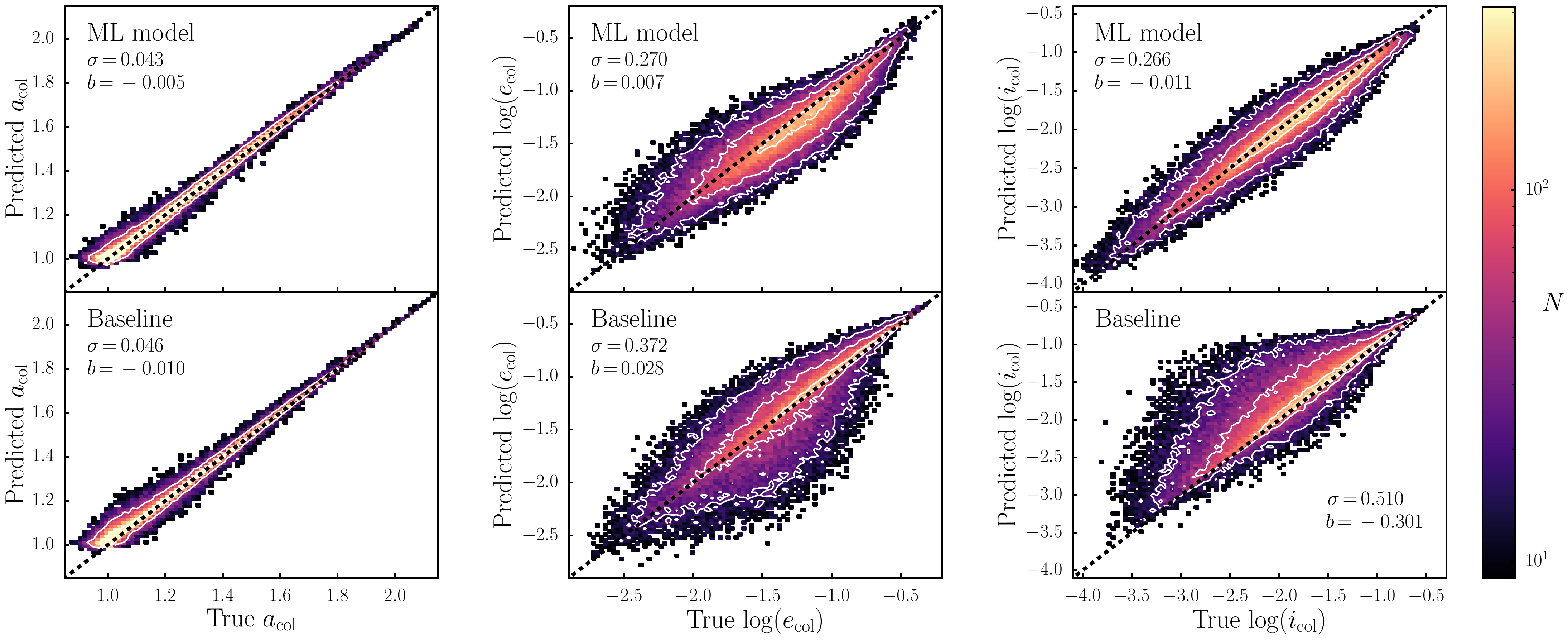}
\caption{Comparison between the performance of the orbital outcome regressor (top row) and a non-ML baseline model (bottom row) on the validation set of $103$,$404$ three-planet systems. Predicted orbital elements (i.e., semi-major axis, eccentricity, and inclination) for the new, merged planets are plotted against their true orbital elements. The ML model predicts orbital elements with somewhat less scatter and bias about the true values than the baseline model, approaching the accuracy limits imposed by chaos (see Table~\ref{table:performance_comparison}).}
\label{fig:regression_comparison1}
\end{figure*}

Combining our collision outcome model with the stability predictions provided by SPOCKII allows us to create a rapid giant impact simulation emulator. We outline our iterative emulator model below.

Beginning with a system of overly packed planets, the giant impact emulator (1) partitions the system into trios of adjacent planets, (2) predicts the instability time of each trio using SPOCKII, and (3) predicts the merger outcome of the trio with the shortest instability time and replaces the planets. This process is repeated until the system is deemed stable by SPOCKII (i.e., the full planetary system has an instability time larger than a user-chosen maximum). In step (2), we determine each trio's instability time using the median of SPOCKII's instability time samples. We find that the emulator's predictions compare much more accurately with $N$-body results (see Section~\ref{sec:planet_formation_preds}) when using median instability time predictions, as opposed to directly sampling SPOCKII's predicted instability time posterior. In step (3), we first predict the probabilities of a collision occurring between planets $1$\,--\,$2$, $2$\,--\,$3$, or $1$\,--\,$3$ in the unstable planet trio using the collision classifier model. Then, we sample the probabilities to determine which pair of planets to combine, and we predict the resulting orbital elements using the orbital outcome regressor model. Results from our giant impact emulator are presented in Section~\ref{sec:planet_formation_preds} and caveats are discussed in Section~\ref{sec:discussion}.

\section{Results}
\label{sec:results}

\subsection{Performance on validation set}
\label{sec:validation_set}

In this section, we study the performance of our ML model, including both collision classifier and orbital outcome regressor parts, on systems from the $103$,$404$ system validation set. Interpreting the performance of our model is complicated by the chaotic nature of multiplanet systems, which can cause systems with indistinguishable initial conditions to rapidly diverge, resulting in different late-time outcomes \citep[e.g.,][]{Rice2018, Hussain&Tamayo2020}. Properly evaluating the performance of our model therefore requires quantifying the limits on prediction accuracy imposed by chaos.

To isolate the influence of chaos, we perform ``shadow integrations'' of $200$ randomly selected systems from the validation set, in which we slightly perturb the initial conditions of the systems and re-integrate it. For each validation system, we construct $250$ shadow realizations where we have perturbed the initial Cartesian positions and velocities of the three planets (in the barycentric frame) by random factors drawn from a Gaussian distribution $\mathcal{N}(\mu\,{=}\,1.0,~\sigma\,{=}\,10^{-12})$. The resulting $250$ shadow systems represent equally valid possible long-term evolutions of the system, allowing us to quantify the variety of potential outcomes.

\subsubsection{Collision classifier}
\label{sec:validation_classifier}

\begin{figure*}
\centering
\includegraphics[width=\textwidth]{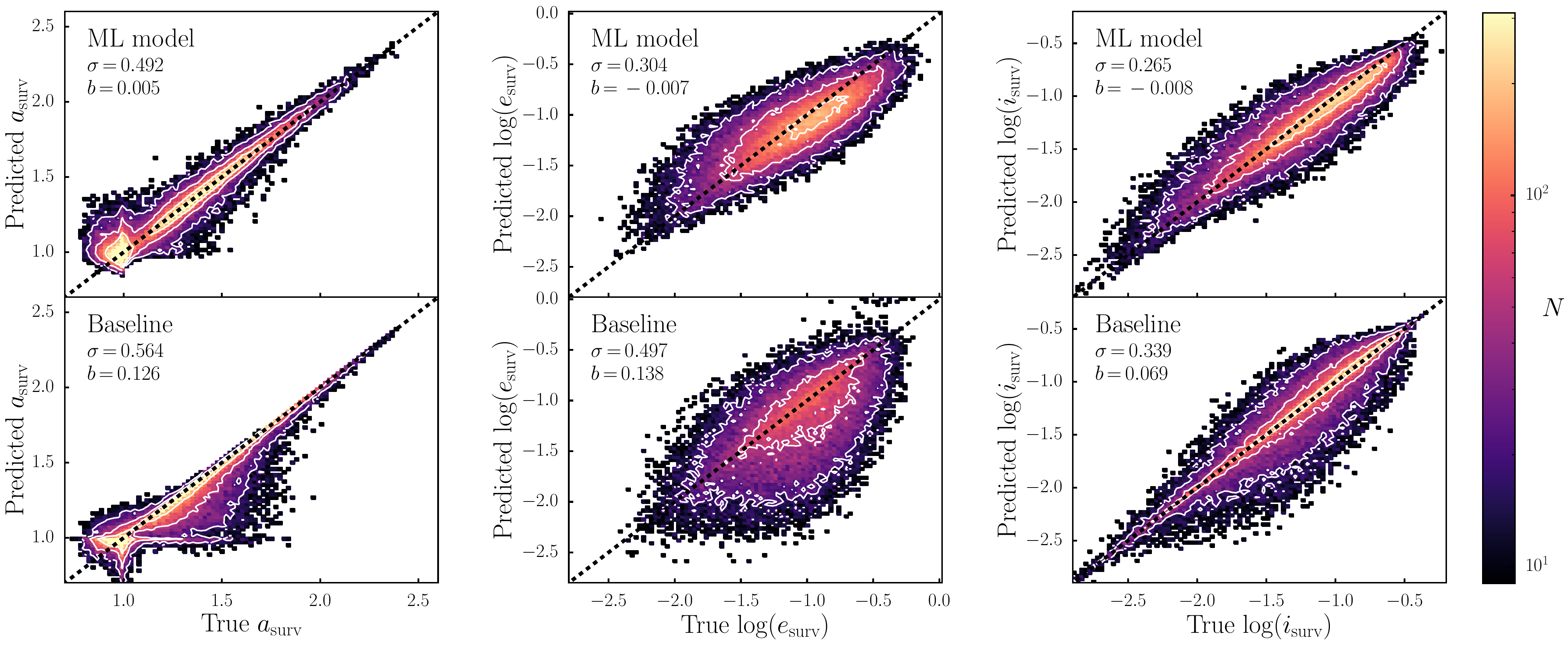}
\caption{Same comparison as Fig.~\ref{fig:regression_comparison1}, now for the three post-collision orbital elements of the surviving planet that is not involved in the merger. The orbital outcome regressor predicts semi-major axis, eccentricity, and inclination with less scatter and bias than the non-ML baseline model, approaching the limit imposed by chaos (see Table~\ref{table:performance_comparison}).}
\label{fig:regression_comparison2}
\end{figure*}

Applying our collision classifier model to the $20$\,\% validation set, we find that the model correctly predicts the planet pair that experiences a collision for $66.5$\,\% of validation systems. This is an improvement over chance ($33.3$\,\% accuracy) and always guessing the most common outcome (collision between planet $1$ and $2$; $43.3$\,\% accuracy), but further interpreting the model's performance requires quantifying the limits on prediction accuracy imposed by chaos.

In many instances, shadow realizations of the same system ultimately experience collisions between a different planet pair. Using our set of shadow integrations, we measure the fractions $f_{\mathrm{planets}\,i-j}$ of shadow systems in which a collision occurs between planets $i$ and $j$, which vary widely depending on the system. In Fig.~\ref{fig:classification_accuracy}, we compare the three collision probabilities predicted by the ML model with the three true collision fractions for each of the $200$ validation systems (we report $\sigma$, the root-mean-squared error, and $b$, the mean of the residuals). Based on a single, short integration of the validation systems, the ML model successfully predicts the probabilities of a collision occurring between the different planet pairs, with a scatter of $\sim\,10$\,\% and very little bias about the true probabilities. Increasing the number of shadow runs does not decrease the scatter, indicating that it is dominated by the error in the ML model's predictions.

To help interpret the performance of the ML model, we considered a variety of non-ML approaches to predicting the collision probabilities. All baseline models we constructed performed poorly, even when tuned to fit the true $f_{\mathrm{planets}\,i-j}$ values (see Appendix~\ref{sec:classifer_baseline}).

\subsubsection{Orbital outcome regressor}
\label{sec:validation_regressor}

Next, we analyze the performance of the orbital outcome regression model. We compare the orbital elements of the newly merged planets, as predicted by the orbital regression model, with the true values for the $20$\,\% validation set in Fig.~\ref{fig:regression_comparison1}. The ML model predictions exhibit little bias about the true values but are otherwise difficult to interpret without a baseline to compare against. The best baseline model we found for predicting the post-collision orbital elements of the merged planet relied on a mass-weighted average of the orbital elements of the two pre-merger planets over the $10^4$ orbit integration. In contrast with the task of predicting collision probabilities, we find that the baseline model predicts post-collision orbital elements relatively well (see bottom panel of Fig.~\ref{fig:regression_comparison1}). Nonetheless, our ML model improves moderately on the scatter and bias of the baseline model, particularly for the inclination of the newly merged planet.

Figure~\ref{fig:regression_comparison2} shows the orbital elements of the planet that is not involved in the collision (the ``surviving planet''), as predicted by our ML-based model and a baseline model, for the $20$\,\% validation set. We create a non-ML baseline model, in this case, by taking the predicted inclination to be the average inclination of the surviving planet over the $10^4$ orbit integration, then we determine the planet's semi-major axis and eccentricity by requiring the conservation of energy $E$ and the $z$-component of angular momentum $L_z$. Note that $E$ is not perfectly conserved during collisions, but the loss of energy during inelastic collisions in our training set simulations is small (${\sim}\,0.05$\,\%), so we may use the approximate conservation of $E$ (and the baseline prediction of $a_\mathrm{col}$) to predict $a_\mathrm{surv}$. On the other hand, all components of $L$ are conserved during collisions, but $L_x$ and $L_y$ depend on $\Omega_\mathrm{col}$ and $\Omega_\mathrm{surv}$, which are unknown, so we leverage only the conservation of $L_z$. Similarly to the orbital elements of the new, merged planet (Fig.~\ref{fig:regression_comparison1}), the ML model predicts the orbital elements of the surviving planet with less scatter and bias than the baseline model.

We may also use the conservation of $L_z$ and the near-conservation of $E$ as a test of the performance of the orbital outcome model. For the set of validation systems, the predicted post-collision states have a median percent error in $L_z$ of $0.3$\,\% and a median percent error in $E$ of $0.6$\,\%. The near-perfect conservation of $L_z$ and $E$ by the model on systems from outside the training set provides additional support for its performance.

\begin{table}
\centering
\caption{Root-mean-squared scatter $\sigma$ and bias $b$ (defined as the mean of residuals) of the orbital outcome regressor and a baseline model on the $103$,$404$ validation systems. For comparison, we report the accuracy limits imposed by chaos (see Section~\ref{sec:validation_set} for more details).}
\begin{tabular}{cccc}
 \hline
 Target & Method & Scatter $\sigma$ & Bias $b$ \\
 \hline
  $a_\mathrm{col}$ & Baseline model & $0.046$ & $-0.010$\\
  $a_\mathrm{col}$ & ML model & $0.043$ & $-0.005$\\
  $a_\mathrm{col}$ & Limit from chaos & $0.040$ & \pp$0.000$\\
 \hline
  $\log(e_\mathrm{col})$ & Baseline model & $0.372$ & \pp$0.028$\\
  $\log(e_\mathrm{col})$ & ML model & $0.270$ & \pp$0.007$\\
  $\log(e_\mathrm{col})$ & Limit from chaos & $0.247$ & \pp$0.000$\\
 \hline
  $\log(i_\mathrm{col})$ & Baseline model & $0.510$ & $-0.301$\\
  $\log(i_\mathrm{col})$ & ML model & $0.266$ & $-0.011$\\
  $\log(i_\mathrm{col})$ & Limit from chaos & $0.242$ & \pp$0.000$\\
 \hline
  $a_\mathrm{surv}$ & Baseline model & $0.564$ & \pp$0.126$\\
  $a_\mathrm{surv}$ & ML model & $0.492$ & \pp$0.005$\\
  $a_\mathrm{surv}$ & Limit from chaos & $0.494$ & \pp$0.000$\\
 \hline
  $\log(e_\mathrm{surv})$ & Baseline model & $0.497$ & \pp$0.138$\\
  $\log(e_\mathrm{surv})$ & ML model & $0.304$ & $-0.007$\\
  $\log(e_\mathrm{surv})$ & Limit from chaos & $0.282$ & \pp$0.000$\\
 \hline
  $\log(i_\mathrm{surv})$ & Baseline model & $0.339$ & \pp$0.069$\\
  $\log(i_\mathrm{surv})$ & ML model & $0.265$ & $-0.008$\\
  $\log(i_\mathrm{surv})$ & Limit from chaos & $0.243$ & \pp$0.000$\\
 \hline
 $E$ & Baseline model & $0.000$ & $0.000$\\
 $E$ & ML model & $2\,{\times}\,10^{-5}$ & $-3\,{\times}\,10^{-6}$\\
 \hline
 $L_z$ & Baseline model & $0.000$ & $0.000$\\
 $L_z$ & ML model & $5\,{\times}\,10^{-6}$ & $-9\,{\times}\,10^{-7}$\\
 \hline
\end{tabular}
\label{table:performance_comparison}
\end{table}

Lastly, using our set of shadow integrations, we quantify the limits on orbital element prediction accuracy imposed by chaos. To do so, for each of the $200$ validation systems, we first bin the shadow integrations according to which two planets experienced a collision. Then, we measure the scatter in the orbital elements about the mean values across the shadow system realizations (which we find to be distributed approximately log-normally), thereby isolating the unavoidable scatter due to the chaotic nature of these simulations. Limits on prediction accuracy due to chaos, along with the performance of the ML and baseline models are reported in Table~\ref{table:performance_comparison}. For all six predicted orbital elements, the accuracy of our ML model is approaching the limit imposed by chaos, indicating that there is little room for improvement. Note, however, that there are likely complex correlations between the outputs of the model.

\subsection{Predicting giant impact outcomes}
\label{sec:planet_formation_preds}

\begin{figure*}
\centering
\includegraphics[width=\textwidth]{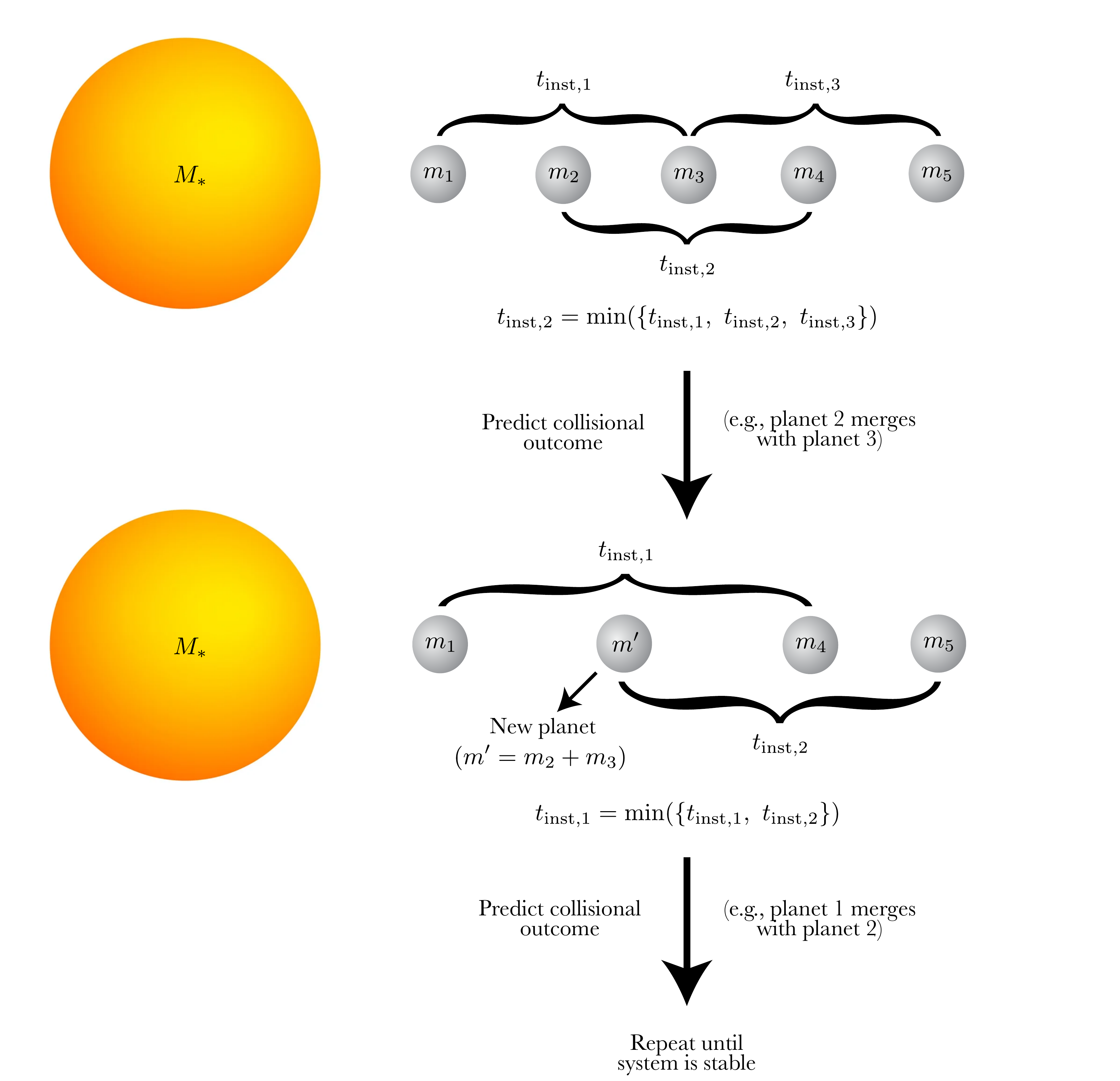}
\caption{Schematic of the iterative giant impact emulator. At each step, instability times for sub-trios of planets are predicted using SPOCKII \citep{Cranmer2021}. Then, with the machine learning model presented in this work, we predict the collisional outcome of the planet trio with the shortest instability time, replace the trio with the two new planets, and repeat until SPOCKII identifies the system as long-term stable.}
\label{fig:process_schematic}
\end{figure*}

\begin{figure*}
\centering
\includegraphics[width=\textwidth]{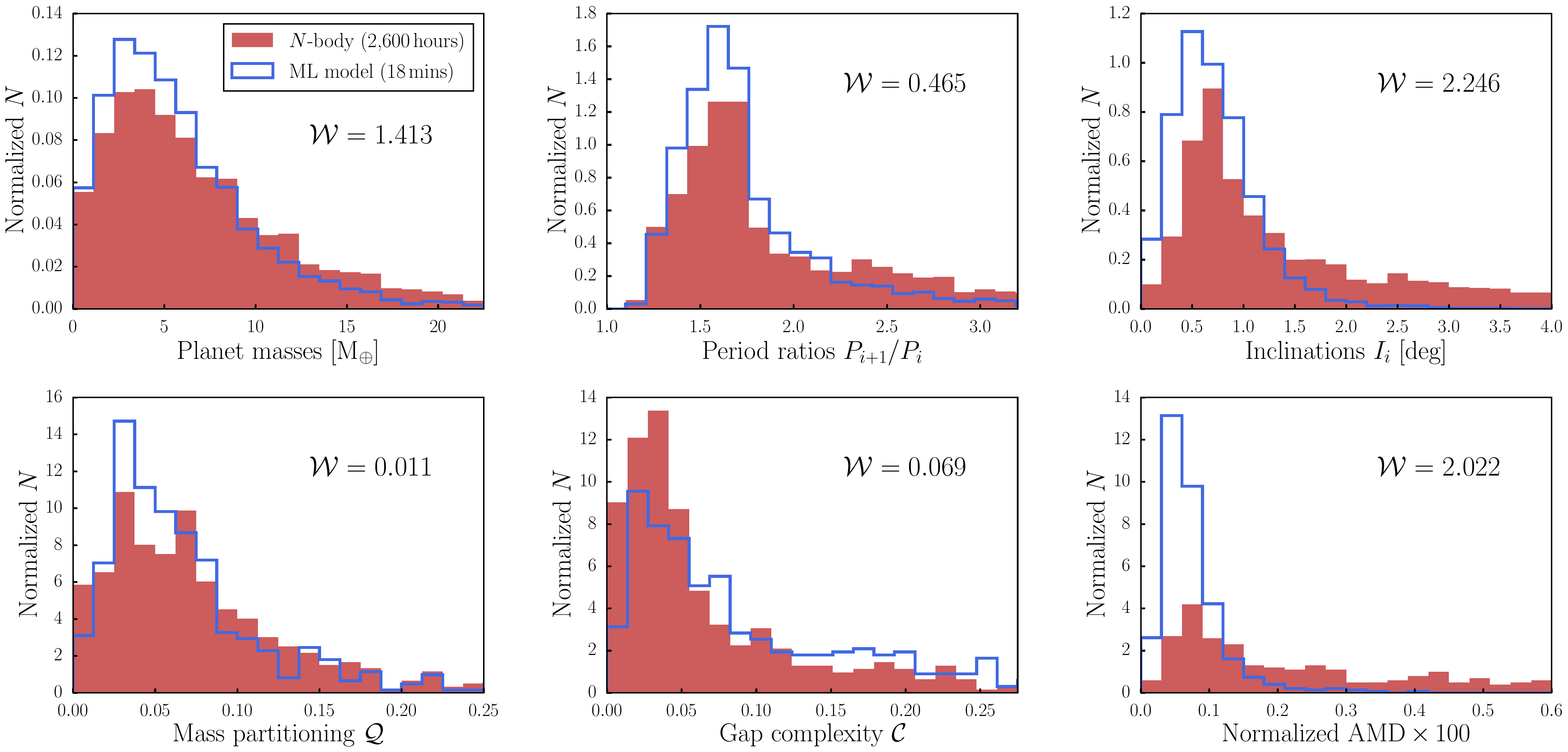}
\caption{Comparison between the properties of $500$ synthetic planetary systems from long-term $N$-body simulations and our ML-based giant impact emulator (which is ${\sim}\,10^4$ faster than full $N$-body simulations). The masses, spacings, and orbital inclinations of the planets in the ML-predicted systems agree well with those from $N$-body simulations (top row). System-level properties of the ML-predicted systems also agree with the $N$-body systems (bottom row), although we find the ML-predicted systems to be somewhat less dynamically excited. The $1D$ Wasserstein distance $\mathcal{W}$ between the two distributions are reported in each panel. Our ML-based model outperforms a non-ML baseline model, whose predictions agree less well with the true $N$-body distributions (for the quantities shown in the top-left to the bottom-right panels, $\mathcal{W}\,{=}\,0.281,\,0.507,\,2.293,\,0.038,\,0.096,\,2.048$).}
\label{fig:formation_preds}
\end{figure*}

By combining our ML model with SPOCKII, which can rapidly predict instability times, we create an iterative giant impact emulator for modelling the planet formation process (see Fig.~\ref{fig:process_schematic} for a schematic that illustrates this model). Below, we evaluate the ability of the emulator to reproduce the results of computationally expensive $N$-body simulations.

For comparison with our giant impact emulator, we carry out $500$ long-term $N$-body simulations of initially overly packed, ten-planet systems. We initialize each system around a solar-mass star, with planet masses drawn randomly from $\mathcal{N}(3M_\oplus,~3M_\oplus)$ and planet radii determined based on the mass-radius relationship from \citet{Wolfgang2016}. Planets were spaced by iteratively drawing initial adjacent-planet period ratios $P_{i+1}/P_i$ from the uniform distribution spanning [$1.10$, $1.75$]. Initial orbital eccentricities and inclinations were drawn from Rayleigh distributions with scale parameters $\sigma\,{=}\,0.01$ and $\sigma\,{=}\,0.5^\circ$, respectively. Initial true longitudes and longitudes of pericenter were drawn uniformly over the range $[0,\,2\pi]$. These initial conditions were chosen to reflect the expected state of multiplanet systems when the gas disk dissipates, and they resemble those adopted when numerically modelling the giant impact process \citep[see, e.g.,][]{Poon2020, Goldberg&Batygin2022, Lammers2023}. Systems were integrated for $10^9$ orbits using the \texttt{MERCURIUS} integrator with a timestep of 1/20th of the innermost planet's initial orbital period, and collisions were treated as perfect inelastic mergers. After $10^9\,P_1$ of dynamical evolution, the systems contain a mean of $5.5$ planets (factor ${\sim}\,2$ reduction), whose orbital properties have been set by the giant impact process.

Figure~\ref{fig:formation_preds} compares the properties of the $500$ planetary systems from $N$-body simulations with the outcomes predicted by our ML-based giant impact emulator, which is $8$,$700$ times faster in this case. The distribution of planet masses, adjacent-planet spacings, and orbital inclinations across the ML-predicted systems agree well with that of the $N$-body systems (for quantitative comparisons, we report the $1D$ Wasserstein distance between the ML and $N$-body distributions in Fig.~\ref{fig:formation_preds}). In addition to the aggregate properties of the planets, the system-level properties of the ML-predicted and $N$-body systems largely agree. To illustrate, we compare the two sets of systems using three metrics: the mass partitioning $\mathcal{Q}$, which describes how mass is distributed among the planets ($\mathcal{Q}\,{=}\,0$ when planets have equal masses and $\mathcal{Q}\,{=}\,1$ when only a single planet is massive; $\mathcal{Q}$ is defined in \citealt{Gilbert&Fabrycky2020}), the gap complexity $\mathcal{C}$, which quantifies the spacing of the planets ($\mathcal{C}\,{=}\,0$ when planets are equally spaced in log-period space; defined in \citealt{Gilbert&Fabrycky2020}), and normalized angular momentum deficit (AMD), which describes the dynamical excitation of the system (see \citealt{Chambers2001, Turrini2020}). Comparing such metrics is a more stringent test of the ML-based emulator because they depend sensitively on which planets are combined (see, e.g., \citealt{Lammers2023}).

The mass partitioning and gap complexity of the ML-predicted systems agree well with that of the $N$-body systems (Fig.~\ref{fig:formation_preds}). However, the normalized AMDs of the ML-predicted system are typically smaller than that of the $N$-body systems, indicating that the ML systems are less dynamically excited (this is also reflected in the distribution of orbital inclinations). This discrepancy is likely caused by the fact that the ML-predicted systems have undergone fewer collisions than the $N$-body systems. Even though SPOCKII predicts the final ML-predicted systems to be stable, they contain a mean of $6.6$ planets, $1.1$ planets above the mean of the true $N$-body systems. We find this trend to hold for other sets of initial conditions as well, so users of the emulator model should be aware that systems tend to be somewhat less dynamically evolved than $N$-body simulations run for the same maximum time. More accurate instability time predictions would help resolve this issue. We note, additionally, that predicting instability times with SPOCKII is the current computational bottleneck of the model --- it is possible to speed up the emulator model (by up to a factor of ${\sim}\,3$) with more rapid instability time predictions.

To help interpret the performance of the ML-based giant impact emulator, and quantify to what extent the strong agreement with $N$-body simulations is due to the tendency of collisions to ``average out,'' we also predict giant impact outcomes for the $500$ initial conditions with a non-ML baseline model. Our baseline model, in this case, relies on the same iterative process as the emulator (Fig.~\ref{fig:process_schematic}), except that at each step, the unstable trio is chosen based on a stability heuristic, the colliding planet pair is determined using our best non-ML baseline model (Appendix~\ref{sec:classifer_baseline}), and mergers are handled with our best non-ML baseline (Figs.~\ref{fig:regression_comparison1},~\ref{fig:regression_comparison2}). We adopt the stability criterion (Eq.~3) from \citet{Tremaine2015}, which attempted to predict the outcomes of the giant impact process analytically. Systems are identified as stable, and the iterative process is stopped, once no planet pairs violate the $10^9$ orbit stability criterion (i.e., all adjacent planet pairs are separated by more than $10$\,mutual Hill radii).

To compare the accuracy of the baseline model with the ML-based emulator, we calculate the $1D$ Wasserstein distance between the properties of the baseline systems and the $N$-body systems. For the properties reported in the top-left to bottom-right panels of Fig.~\ref{fig:formation_preds}, the baseline systems have $\mathcal{W}\,{=}\,0.281,\,0.507,\,2.293,\,0.038,\,0.096,\,2.048$, indicating that the ML-predicted systems more closely match the $N$-body systems for all considered properties except the planet masses. The baseline planet masses agree better with the $N$-body planet masses largely because the multiplicities of the baseline systems (mean of $5.3$ planets) are closer to that of the true $N$-body systems (mean of $5.5$ planets) than the ML-predicted system multiplicities are (mean of $6.6$ planets). Note, however, that the mean number of planets in the baseline systems depends sensitively on the adopted stability heuristic, and despite the similar multiplicities, the other properties of the baseline systems are all less accurate than the ML-predicted systems. 

Comparing with the analytical predictions of \citet{Tremaine2015} is complicated by the presence of the free parameter $\tau$ in their predicted distributions, which depends non-trivially on the set of initial conditions. For normalized eccentricity (Eq.~32 in \citealt{Tremaine2015}), which does not depend on $\tau$, we find the distribution predicted by \citet{Tremaine2015} to significantly overestimate the normalized eccentricities of the $N$-body systems in Fig.~\ref{fig:formation_preds}. Additionally, the absolute eccentricities of the $N$-body systems have a high-eccentricity tail that is not well described by the functional form predicted by \citet{Tremaine2015}, nor a Rayleigh distribution.

As an additional test of the emulator model, we compare the angular momentum $L_z$ and energy $E$ of the final systems with that of the initial conditions. We find that the predicted systems have a median percent error in $L_z$ of just $0.8$\,\% and a median percent error in $E$ of $1.8$\,\%. In practice, the conservation of $L_z$ and $E$ can be used as model diagnostics, allowing the user to reject unphysical systems in which $L_z$ or $E$ are poorly conserved due to hallucinations of the model, which would otherwise be challenging to identify.

\section{Discussion and conclusions}
\label{sec:discussion}

In this work, we present an ML-based approach to predicting the collisional outcomes of multiplanet systems. Breaking this task into two subproblems, we develop two models: a ``collision classifier'' that predicts which pair of planets will collide, and an ``orbital outcome regressor'' that predicts the orbital elements of the two post-collision planets (see Fig.~\ref{fig:NN_schematic}).

Using shadow integrations in which the initial conditions of our synthetic planetary systems are slightly perturbed, we show that our collision classifier model can accurately predict the probabilities of a collision occurring between different planet pairs. With a scatter of ${\sim}\,10$\,\% and little bias about the true collision probabilities, our model significantly outperforms baseline models constructed based on metrics from dynamics theory. Similarly, we find that the orbital outcome regressor outperforms non-ML baselines at the task of predicting the planets' post-collision states. The prediction accuracy of the outcome regressor is only narrowly below the accuracy limits imposed by chaos, and the model's predictions conserve energy and angular momentum to within ${\lesssim}\,1$\,\%.

By combining our model with SPOCKII \citep{Cranmer2021}, which provides rapid instability time predictions, we create an efficient giant impact emulator (see the schematic in Fig.~\ref{fig:process_schematic}). We find that our ML-based emulator can predict the outcomes of $N$-body giant impact simulations with reasonable accuracy and a speed-up of up to four orders of magnitude. Although powerful, this approach requires caution, and we list some notable caveats below.

Firstly, because our ML model was restricted to three-planet systems, instabilities and collisions cannot occur between planet pairs separated by more than one intervening planet (e.g., planets $1$ and $4$). This could present an issue for systems containing bodies that span many orders of magnitude in mass, in which case non-consecutive trios may cause the system to destabilize. Secondly, our model does not account for the scattering of planets that are not included in the unstable planet trio (e.g., if planets $1$, $2$, and $3$ form an unstable trio, a collision between planets $2$ and $3$ could scatter planet $4$). Thirdly, in some instances, the time required for planets to merge after the system has destabilized can be comparable to the system's instability time \citep{Rice2018}, which could result in planets being merged in an incorrect order. Comparisons with $N$-body simulations revealed that there is room for improvement in the accuracy of the giant impact emulator's predictions. Progress will likely require more accurate instability time predictions or changes to the iterative prediction process on which the emulator is based, rather than improvements to the collision classifier and orbital outcome regressor models.

The substantial computational speed-up provided by our collision outcome model and giant impact emulator enables analyses that would not otherwise be feasible, including the ability to carry out more complete explorations of the parameter space of initial conditions. For this reason, we release our model with an easy-to-use API through the SPOCK package,\footnote{\url{https://github.com/dtamayo/spock}} alongside our full training code.\footnote{\url{https://github.com/CalebLammers/ML_for_collisions}}

\section{Acknowledgments} 
\label{sec:acknowledgments}

We thank the anonymous referee for a thorough review that significantly improved the quality of the paper. We would also like to thank Christian Kragh Jespersen, Amir Siraj, Yubo Su, Scott Tremaine, and Joshua Winn for useful discussions. NM acknowledges the support of the Natural Sciences and Engineering Research Council of Canada (NSERC), [RGPIN-2023-04901].

\appendix

\section{Neural network architectures}
\label{sec:architectures}

Here, we describe the architecture of our two MLP models, constructed using the open-source \texttt{PyTorch} \citep{PyTorch} package. Before training the models presented in the main text, the hyperparameters for both models were determined (independently) with Bayesian optimization, as implemented in the \texttt{hyperopt} code. Specifically, for both models, we found the combination of hyperparameters (i.e., the learning rate, weight decay, mini-batch size, number of hidden layers, and number of nodes in the hidden layers) that minimized the loss on a validation set after $10$,$000$ optimization steps. For this, we used a reduced training set of $51$,$702$ systems randomly selected from the full training set, of which $20$\% were reserved for validation.

\subsection{Collision classifier}
\label{sec:classifer_details}

To predict whether a collision will occur between planets $1$\,--\,$2$, planets $2$\,--\,$3$, or planets $1$\,--\,$3$, we use a fully connected MLP with three output nodes. The model includes one hidden layer with $30$ hidden nodes and takes $45$ inputs, corresponding to the three planet masses $\log(m_i)$, along with the mean and standard deviation of $a_i$, $\log(e_i)$, $\log(I_i)$, $\sin(\varpi_i)$, $\cos(\varpi_i)$, $\sin(\Omega_i)$, and $\cos(\Omega_i)$, where $i\,{=}\,1,2,3$. We found that providing the model as input the logarithm of the masses, eccentricities, and inclinations improved the performance of both the classification and regression models, likely because these inputs can span several orders of magnitude.

We use a Rectified Linear Unit (ReLU) activation function on the input and hidden layers and a softmax activation function on the output layer so that the three outputs sum to unity and can be interpreted as probabilities. We train the model using the ADAM optimizer, with a learning rate of $7\,\times\,10^{-4}$ and a weight decay of $1\,\times\,10^{-4}$. The model was trained for $1$,$000$,$000$ optimizer steps using a binary cross-entropy loss function, random mini-batches of size $1$,$000$, and a $20$\% validation set. We explored training the model with other optimizers, including stochastic gradient descent, but found that models trained with ADAM performed best on the validation set.

Before training, we rescale the input masses and orbital elements so that the input features have a mean of zero and a standard deviation of unity across all training systems (during inference, inputs are similarly normalized before being passed to the model). We found that normalizing the input features helped to speed up the training of the model. During training, the model was provided with a random subset of the $100$ orbit samples recorded across the $10^4$ orbit short integration. In detail, we drew $n_t$ uniformly from [$5$, $100$] and calculated the mean and standard deviation of each orbital element from $n_t$ samples selected randomly (without replacement) from the $100$ recorded states. To reflect the varying number of orbit samples and add a small amount of noise to the input features, we draw the means and standard deviations that are passed as input to the MLP model, $\hat{\mu}$ and $\hat{\sigma}$, from their frequentist distributions $\hat{\mu}\,{=}\,\mathcal{N}(\mu,~\sigma/\sqrt{n_t})$ and $\hat{\sigma}\,{=}\,\mathcal{N}(\sigma,~\sigma/\sqrt{2(n_t\,{-}\,1)})$, where $\mu$ and $\sigma$ are the mean and standard deviation measured based on the $n_t$ orbit samples. Varying the number of orbit samples during training acts as a form of data augmentation, which we found improved generalization to the validation set. During inference, the mean and standard deviation are calculated from all $100$ orbital samples with no added statistical noise.

\subsection{Orbital outcome regressor}
\label{sec:regressor_details}

To predict the orbital elements of the two post-collision planets, we use a fully connected multilayer perceptron with six output nodes, corresponding to $a_\mathrm{col}$, $\log(e_\mathrm{col})$, $\log(i_\mathrm{col})$, $a_\mathrm{surv}$, $\log(e_\mathrm{surv})$, and $\log(i_\mathrm{surv})$. The model takes $45$ inputs and consists of a single hidden layer with $60$ hidden nodes. As input, we provide the regression model with the same $45$ values required by the classification model ($\log(m_i)$, $a_i$, $\log(e_i)$, $\log(I_i)$, $\sin(\varpi_i)$, $\cos(\varpi_i)$, $\sin(\Omega_i)$, and $\cos(\Omega_i)$, where $i\,{=}\,1,2,3$), except that the inputs are ordered in order to encode which two planets are to be combined. The orbital properties of the two colliding planets (ordered by semi-major axis) are provided first, followed by those of the surviving planet. We found that re-ordering the inputs to tell the model which two planets are to be combined worked significantly better than passing a separate input value that specified the two colliding planets.

\begin{figure*}
\centering
\includegraphics[width=\textwidth]{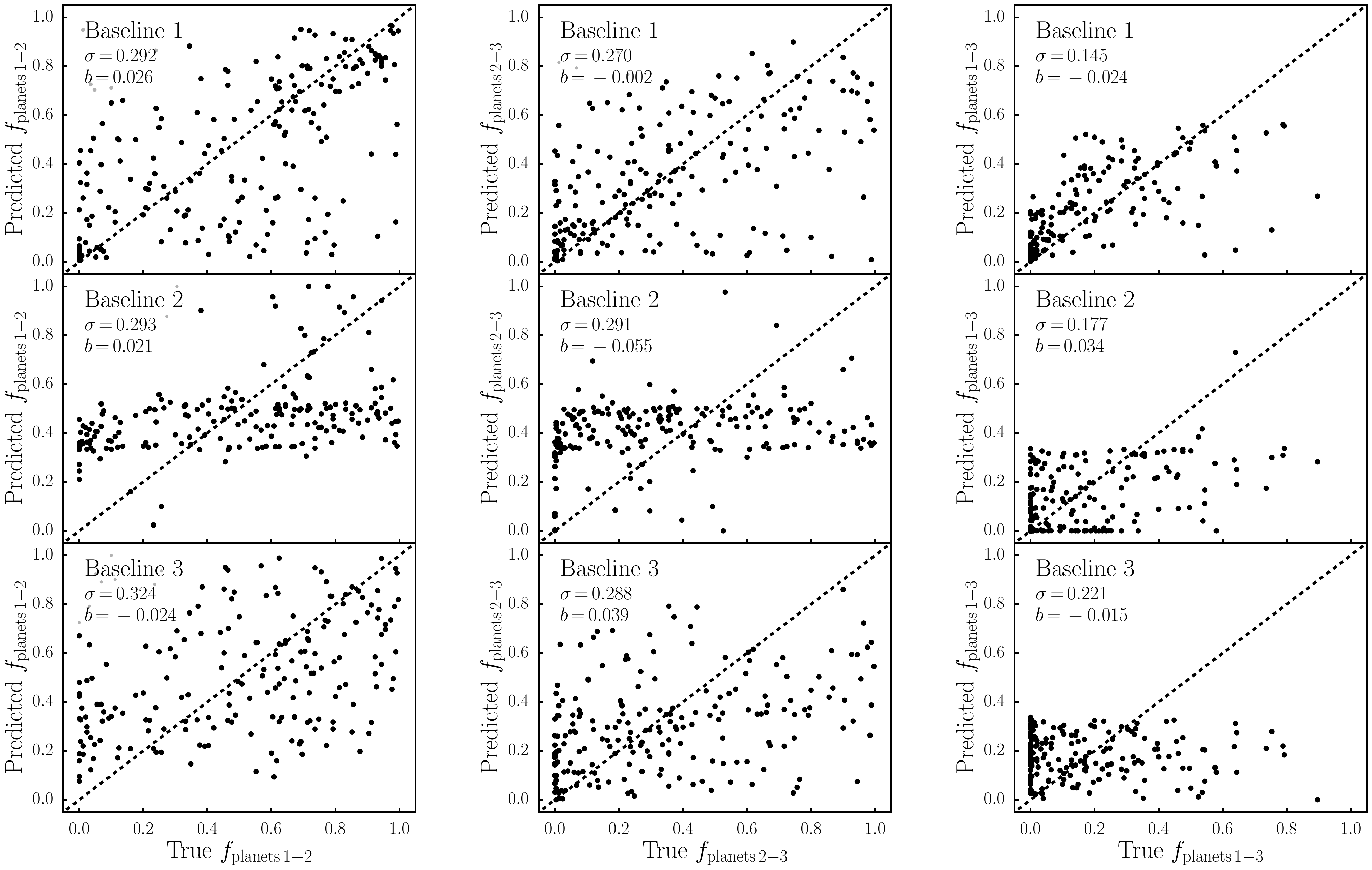}
\caption{Performance of three non-ML baseline classification models on the $200$ validation systems, for comparison with our ML model (Fig.~\ref{fig:classification_accuracy}). From top to bottom, the baseline model predictions rely on the mass ratios of the planets, the amount of angular momentum required for the planets' orbits to cross, and the normalized eccentricity of the planets. The baseline models predict three logits, which are converted to probabilities using a softmax function (see Section~\ref{sec:classifer_baseline}). Despite leveraging dynamics-inspired metrics, the predictions of all three baseline models are poor.}
\label{fig:baseline_classification}
\end{figure*}

For the regression MLP, we use a ReLU activation function on the input and hidden layer, and a linear activation function on the output layer. Similarly to the classification model, we train the model with ADAM (which is, again, the best-performing optimizer) for $1$,$000$,$000$ optimizer steps, in this case with a batch size of $3$,$000$, a learning rate of $7\,\times\,10^{-4}$, and a weight decay of $1\,\times\,10^{-4}$, which were determined based on the hyperparameter optimization process. As with the collision classifier model, during training, we varied the number of orbit samples used to calculate the mean and standard deviation of the orbital elements ($n_t$ was drawn uniformly from [$5$, $100$]), and drew the inputs passed to the model from their corresponding frequentist distributions.

A complication of the orbital element regression problem is that some of the outputs have strict upper limits (e.g., $\log(e_\mathrm{col})\,{<}\,0.0$) and $\log(e_\mathrm{surv})\,{<}\,0.0$). For data points in the training set that are near their upper limit, it is no longer reasonable to assume that their error terms are well-described by symmetric Gaussian distributions, and as a result, the model with minimum mean-squared-error no longer corresponds to the maximum likelihood model. As a result, when trained with a mean-squared-error loss function, we found that the eccentricities predicted by the regression model were systematically biased towards low values. To resolve this, we adopt the loss function $L(y_\mathrm{pred}\,|\,y_\mathrm{true})\,{=}\,\sum_{i\,{=}\,0}^5\,\left((y_{\mathrm{pred},\,i}\,{-}\,y_{\mathrm{true},\,i})^2\,{+}\,\log(1\,{+}\,\mathrm{erf}(y_{\mathrm{max},\,i}\,{-}\,y_{\mathrm{pred},\,i}))\right)$, where $y_\mathrm{pred}$ represents the six outputs of the model, $y_\mathrm{true}$ represents the true values, and $y_\mathrm{max}$ represents the maximum allowed value of each output. The first term in $L(y_\mathrm{pred}\,|\,y_\mathrm{true})$ corresponds to the standard mean-squared-error and the second term accounts for the penalization of predictions near the upper limits (see the Appendix of SPOCKII for a mathematical derivation of the likelihood for data with lower/upper limits). With this loss function, the model predicts post-collision eccentricities with little bias (see the middle panel of Figs.~\ref{fig:regression_comparison1},~\ref{fig:regression_comparison2}). For the interested reader, our full training code is available at \url{https://github.com/CalebLammers/ML_for_collisions}.

\section{Baseline classification model}
\label{sec:classifer_baseline}

Despite extensive theoretical investigation into the destabilization of compact multiplanet systems, there have been relatively few attempts to predict which planets will be involved in the collision. Any binary criterion based on the configuration of the system (e.g., nearest planet pair) inevitably predicts a collision fraction of unity for one planet pair and a fraction of zero for the other two, which is a poor prediction (see Fig.~\ref{fig:classification_accuracy}). Randomly selecting a pair of planets (resulting in $f_{\mathrm{planets}\,i-j}\,{\approx}\,0.33$) is a similarly inadequate prediction. Instead, we approach this task by predicting three logits ($\mathrm{logit}_{1-2}$, $\mathrm{logit}_{2-3}$, and $\mathrm{logit}_{1-3}$) based on the initial conditions of the systems, which are converted to collision fractions according to $f_{\mathrm{planets}\,i-j}\,{=}\,\mathrm{softmax}(\mathrm{logit}_{i-j})$.

We attempted to predict collision probabilities using several different physical properties of the systems. Firstly, motivated by the finding that small planets are preferentially involved in collisions \citep{Lammers2023}, we considered a model that calculates logits using the ratio of the planet masses: $\mathrm{logit}_i\,{=}\,c_i\,\log(\mathrm{max}(m_{i+1}/m_{i},~m_{i}/m_{i+1}))$ where $c_i$ is an arbitrary constant. We tune the three constants $c_i$ to minimize the bias $b$ (defined as the mean of the residuals) with respect to the true collision fractions $f_{\mathrm{planets}\,i-j}$. Secondly, we construct a model in which $\mathrm{logit}_i\,{=}\,c_i\,L_\mathrm{i-j,~cross}$, where $L_\mathrm{i-j,~cross}$ is the minimum orbital momentum that, when transferred to planet $i$ or planet $j$, increases the planets' eccentricity enough to allow for crossing orbits. This criteria thereby takes into account the planets' spacings, masses, and eccentricities. Lastly, we consider a model that calculates logits using the planets' normalized eccentricities (as defined in \citealt{Tamayo2021}), which depend on the planets' spacings, eccentricities, and orbital orientations ($\mathrm{logit}_i\,{=}\,c_i\,|\mathbf{\tilde{e}_{ij}}|$).

The collision probabilities predicted by the three baseline models are compared with the true collision fractions for $200$ validation systems in Fig.~\ref{fig:baseline_classification}. All models perform poorly, with a scatter about the true collision fractions of $\sigma\,{\approx}\,0.3$, three times larger than that of our ML model (Fig.~\ref{fig:classification_accuracy}). Predictions are similarly poor when logits are predicted based on the mean orbital elements over $10^4$ orbits or the state of the system at the end of the $10^4$ orbits, or when using other common stability metrics, like the spacing of the planets in mutual Hill radii. The inability of all baseline models we considered to accurately predict collision probabilities highlights the difficulty of this task.

\bibliography{refs}{}
\bibliographystyle{aasjournal}

\end{document}